\documentclass{aa}
\usepackage{graphicx}
\usepackage{txfonts}
\begin{document}
   \title{The birth rate of supernovae from double-degenerate and core-degenerate systems}


   \author{Xiangcun Meng
          \inst{1,2}
          and
          Wuming Yang \inst{1,3}}

   \offprints{X. Meng}

   \institute{School of Physics and Chemistry, Henan Polytechnic
University, Jiaozuo, 454000, China\\
              \email{xiangcunmeng@hotmail.com}
              \and
Key Laboratory for the Structure and Evolution of Celestial Objects, Chinese Academy of Sciences, Kunming 650011, China
           \and
Department of Astronomy, Beijing Normal University, Beijing
100875, China
             }
   \date{Received; accepted}


  \abstract
   {Some recent observations of the delay-time distribution (DTD) of
   Type Ia supernovae (SNe Ia) seem to uphold the double degenerate (DD) scenario
   as the progenitor model of SNe Ia, but the core-degenerate (CD)
   scenario remains a strong competitor to the DD one.
   }
   {We investigate the effects of metallicity and the different treatments of common
   envelope (CE) on the DTD of SNe Ia by considering the DD and CD scenarios, and check the suggestion that
the total mass of DD system is the main dependent variable of
Phillips relation.}
   {We perform a series of Monte Carlo simulations based on a rapid binary evolution code and
   consider two treatments of CE evolution, i.e.$\alpha$-formalism and $\gamma$-algorithm.}
   {We find that only when the $\alpha$-formalism is considered with a high CE ejection efficiency,
   may the shape of the DTD for DD systems be consistent with that derived observationally,
   i.e. a power law
   of $\sim t^{\rm -1}$, while the value of the birth rate of SNe Ia marginally
   matches observations. For the $\alpha$-formalism with a low CE ejection efficiency
   and the $\gamma$-algorithm, neither the shape of the DTD nor the value of the birth rate
   can be compared with those of the observations. Metallicity may not have a
   significant influence on the shape of DTD, but a lower metallicity may lead to a slightly higher birth rate of SNe Ia by a factor of 2, especially for
   SNe Ia with long delay times. If the results for the single degenerate (SD) channel
   are incorporated into those for the DTD, both the shape of DTD
   and its value may be closely consistent with observations for SNe Ia younger than 2.5 Gyr, and
   SD and DD channels provide comparable contributions to the total SNe Ia,
   while for SNe Ia with delay times longer than 2.5 Gyr, DD is the dominant channel and the birth rate is lower than that derived from
   observations by a factor up to about four. In addition, we calculate the
   evolutions of various integral parameters of DD systems,
   and do not find any one suitable to explain the correlation between the brightness of
   SNe Ia and its delay time. Moreover, there are three channels producing core-degenerate(CD) systems that may contribute a few SNe Ia,
   but the contribution of CD systems
   to the total SNe Ia is no more than 1\%.}
   {There may be other channels or mechanisms contributing to SNe Ia with long delay times.}

   \authorrunning{Meng \& Yang}
   \titlerunning{The birth rate of supernovae from DD and CD systems}
   \maketitle{}
%

\section{Introduction}\label{sect:1}
Type Ia supernovae (SNe Ia) have been found to be important to
many astrophysical fields, especially as principal distance
indicators to measure cosmological parameters, where their use
resulted in the discovery of the accelerating expansion of the
Universe (Riess et al. \cite{REI98}; Schmidt et al.
\cite{SCHMIDT98}; Perlmutter et al. \cite{PER99}). This result was
extremely exciting and implied that dark energy exists. At
present, SNe Ia are regarded as critical cosmological probes for
testing both the evolution of the dark energy equation of state
with time and the evolutionary history of the universe (Riess et
al. \cite{RIESS07}; Kuznetsova et al. \cite{KUZNETSOVA08}; Howell
et al. \cite{HOWEL09}).

However, the precise nature of SNe Ia remains unclear, especially
their progenitor (Hillebrandt \& Niemeyer \cite{HN00}; Leibundgut
\cite{LEI00}; Parthasarathy et al. \cite{PAR07}; Wang \& Han
\cite{WANGB12}). There is a consensus that SNe Ia result from the
thermonuclear explosion of a carbon-oxygen (CO) white dwarf (WD)
in a binary system (Hoyle \& Fowler \cite{HF60}). According to the
nature of the companions of the mass-accreting WDs, two possible
progenitors of SNe Ia have been discussed over the past three
decades, i.e. the single degenerate (SD) model where the companion
is a normal star (Whelan \& Iben \cite{WI73}; Nomoto et al.
\cite{NTY84}), and the double degenerate (DD) model where a CO WD
merges with another CO WD (Iben \& Tutukov \cite{IBE84}; Webbink
\cite{WEB84}). At present, both models cannot be ruled out
completely by observations (see the review by Howell
\cite{HOWEL11}).

To distinguish between these different progenitor models, it is
important to measuring the delay-time distribution (DTD), where
the delay time is the elapsed time between the primordial system
formation and the explosion as a supernova (SN) event and the DTD
is the SN birth rate versus their delay time for a single
starburst. An increasing amount of observational studies of the
DTD of SNe Ia have found that the DTD follows a power-law form of
$t^{\rm -1}$ (Totani et al. \cite{TOTANI08}; Maoz \& Mannucci
\cite{MAOZ11}), which is difficult to interpret in terms of the SD
model (Meng \& Yang \cite{MENGYANG10a}), but in contrast can be
naturally explained by the DD model (Yungelson \& Livio
\cite{YUN00}; Mennekens et al. \cite{Mennekens10}). Some
theoretical and observational studies have shown that metallicity
may have a strong influence on the birth rate of SNe Ia (Khan et
al. \cite{KHAN11}; Meng et al. \cite{MENGXC11}). We wish to
establish the effect of metallicity on the birth rate of SNe Ia in
the DD scenario.

When SNe Ia are used as distance indicators, the Phillips relation
is adopted, which is a linear relation between the absolute
magnitude of SNe Ia at maximum light and the magnitude drop in the
B band light curve during the first 15 days following the maximum
light (Phillips \cite{PHI93}). This relation was motivated by the
observations of two peculiar SN events (SN 1991bg and SN 1991T)
and implies that the brightness of SNe Ia is determined mainly by
one parameter. The amount of $^{\rm 56}$Ni formed during the SN
explosion dominates the maximum luminosity of SNe Ia (Arnett
\cite{ARN82}). However, the origin of the variation in the amount
of $^{\rm 56}$Ni for different SNe Ia is still unclear
(Podsiadlowski et al. \cite{POD08}). Observationally, the most
luminous SNe Ia always occur in spiral galaxies, while both spiral
and elliptical galaxies host systematically dimmer SNe Ia, which
leads to a dimmer mean peak SN brightness in elliptical than
spiral galaxies (Hamuy et al. \cite{HAM96}; Brandt et al.
\cite{BRANDT10}). In addition, the mean peak brightness of SNe Ia
in a galaxy varies less in the outer than the inner regions (Wang
et al. \cite{WAN97}; Riess et al. \cite{RIE99}). In other words,
age could be the most important factor in determining the
luminosity of SNe Ia, and dimmer SNe Ia have a wide age
distribution (Gallagher et al. \cite{GALLAGHER08}; Neill
\cite{NEILL09}; Howell et al. \cite{HOWEL09b}). From these
observations, we concluded that the range and average value of the
brightness of SNe Ia decrease with their delay time. If the
maximum luminosity of SNe Ia is mainly determined by one parameter
as shown by the Phillips relation, we may expect that the range of
the parameter decreases, and its average value either increases or
decreases with the age of SNe Ia.

Many efforts have been made to resolve this problem. Some
multi-dimensional numerical simulations have shown that varying
the ignition intensity at the center of WDs or the transition
density from deflagration to detonation might enable us to
determine the underlying cause of the observed correlation between
peak luminosity and light-curve width (Hillebrandt \& Niemeyer
\cite{HN00}; H\"{o}flich et al. \cite{HOFLIC06, HOFLIC10}; Kasen
et al. \cite{KASEN10}). In addition, the ratio of
nuclear-statistical-equilibrium to intermediate-mass elements in
the explosion ejecta may be a key parameter in determining the
width of SN Ia light curve and its peak luminosity (Pinto \&
Eastman \cite{PINTO01}; Mazzali et al.
\cite{MAZZALI01,MAZZALI07}). Lesaffre et al. (\cite{LESAFFRE06})
suggested that the central density of the WD at ignition may be
the origin of the Phillips relation in a systematic study of the
sensitivity of ignition conditions for H-rich Chandrasekhar-mass
single-degenerate exploders to various properties of the
progenitors (see also Podsiadlowski et al. \cite{POD08}), which
was upheld by detailed multi-dimensional numerical simulations of
explosion (Krueger et al. \cite{KRUEGER10}). Moreover, metallicity
may also affect the production of $^{\rm 56}$Ni, thus the maximum
luminosity, both in theory (Timmes et al. \cite{TIM03}; Travaglio
et al. \cite{TRA05}; Podsiadlowski et al. \cite{POD06}; Bravo et
al. \cite{BRAVO10}) and observations (Branch \& Bergh \cite{BB93};
Hamuy et al. \cite{HAM96}; Wang et al. \cite{WAN97}; Cappellaro et
al. \cite{CAP97}; Shanks et al. \cite{SHA02}). On the basis of the
suggestion that the ratio of the carbon to oxygen (C/O) of a white
dwarf at the moment of explosion could be the dominant parameter
determining the production of $^{\rm 56}$Ni (Nomoto et al.
\cite{NOM99, NOM03}), Meng \& Yang (\cite{MENGYANG11a}) found that
the effects of metallicity and C/O on the production of $^{\rm
56}$Ni may complement each other. Most of these discussions above
focused on the Chandrasekhar mass model, in which the WDs explode
as SNe Ia when their masses are close to the Chandrasekhar mass
limit. However, the sub-Chandrasekhar mass model may still be able
to explain SNe Ia such as 1991bg, as well as normal SNe Ia (Sim et
al. \cite{SIM10}; Ruiter et al. \cite{RUITER11}). In addition,
observations of several very bright SNe Ia implied that their
progenitor WDs might have a super-Chandrasekhar mass (Astier et
al. \cite{ASTIER06}; Howell et al. \cite{HOW06}; Hicken et al.
\cite{HICKEN07}; Scalzo et al. \cite{SCALZO10}; Yuan et al.
\cite{YUAN10}; Tanaka et al. \cite{TANAKA10}; Yamanaka et al.
\cite{YAMANAKA10}). Although these very bright SNe Ia may also be
explained in terms of the SD scenario (Chen \& Li \cite{CHENWC09};
Liu et al. \cite{LIUWM10}; Hachisu et al. \cite{HKHN12}), that the
DD systems are the progenitor of the very bright SNe Ia still
cannot be completely ruled out at present. Howell (\cite{HOWEL11})
suggested that the DD scenario has a natural explanation of the
higher SN luminosity in young environments, i.e. younger, more
massive stars produce more massive white dwarfs that have more
potential fuel than less massive mergers, and then produce a
bright SNe Ia (see also Maoz \& Mannucci \cite{MAOZ11}). This
suggestion still needs to be checked carefully.

However, earlier numerical simulations showed that the most
probable fate of the DD merger is an accretion-induced collapse
(AIC) and, finally,  neutron star formation (see the review by
Hillebrandt \& Niemeyer \cite{HN00}). Even if the merger survived
the AIC, a super wind from giant-like structure before supernova
explosion would occur and the remnant might lose about 0.5
$M_{\odot}$ and shrink in mass to below the critical mass for
explosion (Soker \cite{SOKER11}). Following the suggestion of
Sparks \& Stecher (\cite{SPARKS74}) and Livio \& Riess
(\cite{LR03}), Kashi \& Soker (\cite{KASHI11}) developed a
core-degenerate (CD) model to overcome the drawbacks of the DD
model, in which a white dwarf merges with the core of an
asymptotic giant branch (AGB) star shortly after a common envelope
(CE) phase. However, they did not consider the evolution of the
birth rate from the CD supernova. In this paper, we study the
evolution in terms of a detailed binary population synthesis (BPS)
study.

In Sect. \ref{sect:2}, we simply describe our method, and present
the calculation results in Sects. \ref{sect:3} and \ref{sect:4}.
In Sect. \ref{sect:5}, we show our discussions and main
conclusions.


\section{METHOD}\label{sect:2}

\subsection{Common envelope}\label{subs:2.1}
The CE phase is very important to the formation of DD systems.
During binary evolution, the mass ratio ($q=M_{\rm donor}/M_{\rm
accretor}$) is a crucial parameter. If it is larger than a
critical mass ratio, $q_{\rm c}$, mass transfer between the two
components is dynamically unstable and a CE forms
(Paczy$\acute{\rm n}$ski \cite{PAC76}). The ratio $q_{\rm c}$
varies with the evolutionary state of the donor star at the onset
of Roche lobe overflow (RLOF) (Hjellming \& Webbink\cite{HW87};
Webbink \cite{WEBBINK88}; Han et al. \cite{HAN02}; Podsiadlowski
et al. \cite{POD02}; Chen \& Han \cite{CHE08}). In this study, we
adopted $q_{\rm c}$ = 4.0 when the donor star is on the main
sequence (MS) or crossing the Hertzsprung gap (HG). This value is
supported by detailed binary evolution studies (Han et al.
\cite{HAN00}; Chen \& Han \cite{CHE02, CHE03}). If the donor star
is on either the first giant branch (FGB) or the AGB, we use
\begin{equation} q_{\rm c}=\left[1.67-x+2\left(\frac{M_{\rm c}}{M}\right)^{\rm 5}\right]/2.13,  \label{eq:qc}
  \end{equation}
where $M_{\rm c}$ is the core mass of the donor star, and $x={\rm
d}\ln R_{\rm 1}/{\rm d}\ln M$ is the mass--radius exponent of the
donor star, which varies with composition. If the mass donors are
naked helium giants, then $q_{\rm c}$ = 0.748 based on Eq.
(\ref{eq:qc}) (see Hurley et al. \cite{HUR02} for details).

Embedded in the CE are the dense core of the donor star and the
secondary. Owing to frictional drag with the envelope, the orbit
of the embedded binary decays, and a large part of the orbital
energy released in the spiral-in process is injected into the
envelope (Livio \& Soker \cite{LS88}).

The CE evolution is very complicated and different authors may use
different methods to treat it in their BPS studies. There are two
dominant methods for treating the CE evolution, i.e. the
$\alpha$-formalism ensuring energy balance and the
$\gamma$-algorithm ensuring angular momentum balance. The
$\alpha$-formalism, which is widely used, may closely reproduce
the orbital-period distribution of WD + MS systems as noted by
Zorotovic et al. (\cite{ZOROTOVIC10}) (see also Hurley et al.
\cite{HUR02} and Webbink \cite{WEBBINK07}), while it may be unable
to describe the production of a close pair of white dwarfs. To
solve this problem, Nelemans et al. (\cite{NELEMANS00}) and
Nelemans \& Tout (\cite{NELEMANS05}) developed the
$\gamma$-algorithm, which may explain the formation of all kinds
of close binaries. In this paper, both of these descriptions are
applied.

For the $\alpha$-formalism, the final orbital separation $a_{\rm
f}$ after CE phase is obtained by
\begin{equation}
\frac{a_{\rm f}}{a_{\rm i}}=\frac{M_{\rm
c}}{M}\left(1+\frac{2M_{\rm e}a_{\rm i}}{\alpha_{\rm CE}\lambda
mR_{\rm 1}}\right)^{\rm -1},
  \end{equation}
where $a_{\rm i}$ is the initial orbital separation at the onset
of the CE,  the masses $M$, $M_{\rm c}$, and $M_{\rm e}$ are those
of the donor, the donor core, and the envelope , respectively,
$R_{\rm 1}$ is the radius of the donor, and $m$ is the companion
mass. The parameter $\alpha_{\rm CE}$ is the CE ejection
efficiency, i.e. the fraction of the released orbital energy used
to eject the CE, and $\lambda$ is a structure parameter relying on
the evolutionary stage of the donor. At present, the value of
$\alpha_{\rm CE}$ is very uncertain and may vary from 0.4 to 3.0
(see the review of Ivanova \cite{IVANOVA11}). Because the thermal
energy in the envelope is not incorporated into the binding energy
in this paper, $\alpha_{\rm CE}$ may be greater than 1 (see Han et
al. \cite{HAN95} for details about the thermal energy). In this
paper, we set $\alpha_{\rm CE}$ to either 1.0 or 3.0 when studying
the DD and CD scenarios, and $\alpha_{\rm CE}=1.5$ or $2.0$ were
also tested. For $\lambda$, we assumed that it is a constant
($\lambda=0.5$, de Kool, van den Heuvel \& Pylyser
\cite{DEKOOL87}).

In the $\gamma$-algorithm, $a_{\rm f}$ was obtained from
\begin{equation}
\frac{a_{\rm f}}{a_{\rm i}}=\left(\frac{M}{M_{\rm c}}\right)^{\rm
2}\left(\frac{M_{\rm c}+m}{M+m}\right)\left(1-\gamma_{\rm
CE}\frac{M_{\rm e}}{M+m}\right)^{\rm 2}. \label{gamma}
  \end{equation}
where $\gamma_{\rm CE}$ is a free parameter. Based on the results
of Nelemans \& Tout (\cite{NELEMANS05}), we set $\gamma_{\rm
CE}=1.5$.

\subsection{Basic parameters of our Monte Carlo simulations}\label{subs:2.2}
To investigate the birthrate of SNe Ia in DD systems, we followed
the evolution of $1\times10^{\rm 7}$ binaries via Hurley's rapid
binary evolution code (Hurley et al. \cite{HUR00, HUR02}). The
primordial binary samples are generated in a Monte Carlo way and a
circular orbit was assumed for all binaries. The basic parameters
for the simulations were as follows: (1) a constant star formation
rate (SFR) of $5$ $M_{\odot}$ ${\rm yr^{-1}}$ over the past 15 Gyr
or a single starburst of $10^{\rm 11} M_{\odot}$; (2) the initial
mass function (IMF) of Miller \& Scalo (\cite{MIL79}); (3) a
constant mass-ratio distribution; (4)all stars are members of
binary systems and the distribution of separations is constant in
$\log a$ for wide binaries, where $a$ is the orbital separation,
and falls off smoothly at small separations, where $a=10R_{\odot}$
is the boundary for wide and close binaries; (5) metallicities
were chosen to be $Z=0.02$, $0.002$, and 0.0001 (see Meng \& Yang
\cite{MENGYANG10a} for details of these input parameters).

We tracked the evolutions of the $1\times10^{\rm 7}$ binaries
until they become DD systems. Following the DD system,
gravitational wave radiation (GWR) dominates the evolution of the
system on a timescale $t_{\rm GW}$ (Landau \& Lifshitz
\cite{LANDAU62})
\begin{equation}
t_{\rm GW}{\rm (yr)}= 8\times10^{\rm 7}\times\frac{(M_{\rm
1}+M_{\rm 2})^{\rm 1/3}}{M_{\rm 1}M_{\rm 2}}P^{8/3},\label{eq:gw}
  \end{equation}
where $P$ is the orbital period of the DD system in hours, and
$M_{\rm 1}$ and $M_{\rm 2}$ are the masses of the two WDs in solar
masses, respectively. The time elapsed from the birth of
primordial binary system to the occurrence of SN Ia was then
assumed to be equal to the sum of the timescale on which the
primordial secondary star becomes a WD and the orbital decay time.
We assume that if the total mass of a DD system is $M_{\rm
t}=M_{\rm 1}+M_{\rm 2}\geq1.378M_{\odot}$ and the elapsed time is
shoter than $15$ Gyr, a SN Ia is produced.

\subsection{Evolution channel for the CD system}\label{subs:2.3}
A system consisting of a WD and an AGB core survives a CE phase
and then merges shortly after. The core is more massive than the
WD and when the merging process starts, it is still very hot and
has a larger radius than the cold WD. Hence, the core may be
destroyed to form an accretion disk around the WD. Since the less
massive WD has a shallower gravitational potential, the
temperature in the accretion disk does not reach the ignition
temperature and AIC can be avoided. This scenario is only
applicable when the merging process occurs within $\sim10^{\rm 5}$
yr of the CE phase. The merger is unlikely to explode soon until a
large part of the angular momentum is lost by means of the
magneto-dipole radiation torque (see Ilkov \& Soker \cite{ILKOV11}
for details). Thus, as mentioned in Soker (\cite{SOKER11}), three
key factors distinguishing the CD scenario from the DD channel are
that in the former scenario: (1) the hot core is more massive than
its cold companion WD, (2) the CD system will merge within
$\sim10^{\rm 5}$ yr after the CD system forms\footnote{Note that
the time here is the GWR timescale after the CD system forms and
is not the delay time of a SN Ia.}, and (3) its delay time is
mainly due to the spinning-down time of the merger product, which
results from a magneto-dipole radiation torque (Ilkov \& Soker
\cite{ILKOV11}). Ilkov \& Soker (\cite{ILKOV11}) and Soker
(\cite{SOKER11}) only introduced one channel to produce CD
systems. Three channels could actually produce CD systems,
fulfilling the above criteria. However, different treatments of CE
may trigger different channels as follows. In the following
description, the \emph{primary} is the more massive star in a
primordial binary system, while the \emph{secondary} is the less
massive one.

Case 1: (wind + RL + CE) This channel can be encountered by both
$\alpha$-formalism and $\gamma$-algorithm, but is more common for
$\gamma$-algorithm, and is similar to that described in Soker
(\cite{SOKER11}) and Ilkov \& Soker (\cite{ILKOV11}), which
includes two sub-channels. For sub-channel A, the primordial
zero-age main sequence (ZAMS) mass of a primary is in the range
from 4.0 $M_{\odot}$ to 5.0 $M_{\odot}$, and the mass ratio
($m_{\rm 2}/m_{\rm 1}$) is relatively high(0.85 - 0.95). The
primordial binary system has a very wide separation (wider than
2300 $R_{\odot}$), which permits the primary to evolve into the
thermal pulsing asymptotic giant branch (TPAGB). Before the
primary fills its Roche lobe, it loses a lot of material by its
wind, which results in the following stable RLOF\footnote{The
treatment of RLOF in the Hurley's rapid binary evolution code is a
substantially revised version of that presented by Tout et al.
(\cite{TOUT97}) and the radius-mass exponent $\zeta$ defined by
Webbink (\cite{WEBBINK85}) is used to describe the stability of
mass transfer.}. At this stage, the secondary is still a MS star.
The system becomes a CO WD + MS system after the RLOF. The MS
secondary then has a mass higher than the primordial mass of the
primary. The WD + MS system continues to evolve and the MS
secondary fills its Roche lobe when it becomes a AGB star. Because
of the large mass ratio at this stage, the system enters a CE
phase. If the CE can be ejected, a CD system forms and merges
within $\sim10^{\rm 6}$ yr.

For sub-channel B, the primordial primary usually has a mass of
$3.0-3.5 M_{\odot}$, and the mass ratio is higher than 0.99.
However, the primordial separation for sub-channel B is not as
wide as that for the sub-channel A ($1800 R_{\odot}$ - $2100
R_{\odot}$). The evolutionary path of the sub-channel B is much
similar to that of the sub-channel A and the difference is that
when the primary becomes an AGB star, the secondary is a
horizontal branch (HB) star (central helium burning), not a MS
star.

Case 2: (RL + 4CE) This channel is only encountered when
$\gamma$-algorithm is adopted. The primordial primary has a mass
higher than 6.5 $M_{\odot}$ and the mass ratio is larger than
0.94, even close to 1, but the primordial separation is only about
100 $R_{\odot}$. Owing to the small separation, when the primary
crosses the HG, it fills its Roche lobe and a stable RLOF occurs,
which results in the secondary being more massive than the
primordial primary. The separation of the components then
increases during the stable mass transfer, the primary continues
to evolve to the FGB, and the mass transfer then becomes unstable,
leading to the first CE. After the CE ejection, the system
consists of a naked helium star and a MS star. According to the
$\gamma$-algorithm, the separation does not shrink greatly after
the first CE ejection. Shortly afterwards, the MS star evolves to
the FGB and fills its Roche lobe where the mass transfer is
dynamically unstable, and the system then enters into the second
CE phase. After the CE ejection, the secondary also becomes a
naked helium star, and the system consists of two helium stars.
The mass of the helium star from the secondary is higher than that
of the primary because the secondary at the MS stage is more
massive than the primordial primary for prior stable RLOF
($m_{\rm1}\in[1.22,1.35]M_{\odot}$ and
$m_{\rm2}\in[1.4,1.6]M_{\odot}$, respectively). The helium star
from the primary firstly exhausts its central helium and then
begins to cross the helium HG. If it fills its Roche lobe, the
third CE is expected for a large mass ratio. However, the mass of
the third CE is low compared to the two prior CEs, and the third
CE evolution cannot shrink the separation significantly. The CE is
again ejected and then the primary becomes a CO WD. Helium is also
finally exhausted at the center of the second helium star. As the
second helium star crosses the helium HG, it also fills its Roche
lobe, leading to the fourth CE. After the ejection of the last CE,
a CD system forms. Because of the four CE evolutions, the
separation of the CD system is so close that the system merges
within $\sim10^{\rm 6}$ yr.

Case 3: (RL + CE + RL/wind + CE) This channel is encountered when
an $\alpha$-formalism is adopted. The primordial primary has a
mass of $4.3-5.2 M_{\odot}$ and its mass ratio is lower than in
the above two cases (0.68 - 0.78). The range of primordial
separations is from 20 $R_{\odot}$ to 55 $R_{\odot}$. Because of
the small separation, the primary fills its Roche lobe as it
crosses the HG, where a stable mass transfer occurs. The primary
loses its envelope and becomes a naked helium star. At the same
time, the secondary accretes a lot of material, becoming more
massive than primordial primary. Owing to the stable RLOF, the
separation greatly increases, and the secondary may then evolve to
its HG/RGB phase and fill its Roche lobe. The subsequent mass
transfer is dynamically unstable and a CE forms. After the CE
ejection, the system consists of two naked helium stars, and owing
to the prior stable RLOF, the helium star evolving from the
secondary is more massive ($m_{\rm1}\in[0.7,0.85]M_{\odot}$ and
$m_{\rm2}\in[1.25,1.6]M_{\odot}$, respectively). The helium star
that evolves from the primary firstly exhausts its central helium
and begins to cross the helium HG. At this stage, if the helium
star fills its Roche lobe, a stable RLOF occurs and the helium
star transfers its envelope onto the second helium star and then
becomes a CO WD. It otherwise loses its envelope in terms of a
wind to become a WD. The second helium star also finally exhausts
its central helium, leading to the expansion of its envelope. Its
Roche lobe is then filled and the system enters into the second CE
phase for unstable mass transfer. After the CE ejection, a CD
system forms and merges within $10^{\rm 6}$ yr.

In this paper, we assumed that (1) if a WD produced from a
primordial secondary is more massive than its companion WD, the
system is a potential CD system; (2) if the potential CD system
merges after the emission of GWR within a few$\times10^{5}$ yr,
the merger will explode as a SNe Ia. In addition, to study the
birth rate of SNe Ia from CD systems, we traced the evolution of
$1\times10^{\rm 8}$ binaries. To check the influence of the merger
timescale by GWR on the birth rate of SNe Ia in the CD system, we
considered the cutting time as a free parameter (see Sect.
\ref{sect:4}).

There is actually a ``pollution'' channel, that also fulfills our
criterion (1) for the potential CD systems. In the channel, the
more massive WD also evolves from secondary, but forms earlier
than the second WD. The primordial primary has a mass of around $5
M_{\odot}$, and the mass ratio is close to 1. The primordial
separation is in the range of $130-160 R_{\odot}$, which permits
the primary to evolve to the FGB. A dynamically unstable mass
transfer then results in a CE phase. Following the CE ejection,
the system consists of a naked helium star and a MS one. On the
basis of the $\gamma$-algorithm for CE evolution, the orbital
separation does not shrink greatly after the first CE evolution,
which permits the secondary to evolve to the early AGB phase, and
then the second CE. After the CE is ejected, the secondary becomes
a WD and the separation shrinks greatly. After the helium star
exhausts its central helium, the helium star also fills its Roche
lobe, and the system enters into the third CE phase. After the CE
ejection, the system consists of a less massive hot core and a
more massive cold WD. The following GWR timescale for the channel
may be as long as 1.5 Gyr.

    \begin{figure}
    \centering
    \includegraphics[width=60mm,height=80mm,angle=270.0]{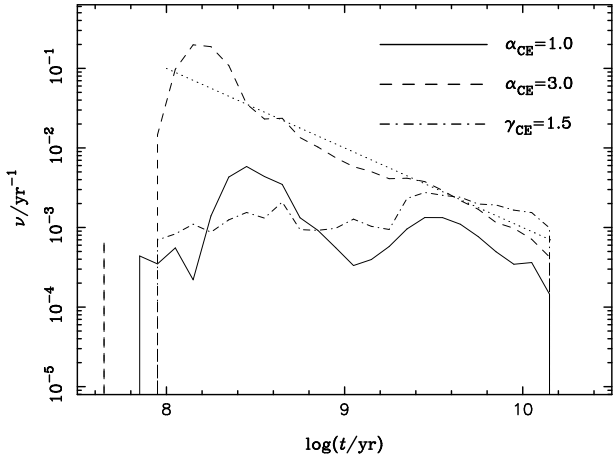}
    \caption{Evolution of the birthrates of SNe Ia for a single
             starburst of $10^{\rm 11}M_{\odot}$, where $Z=0.02$. Solid,
             dashed, and dot-dashed lines show the cases with $\alpha_{\rm CE}=1.0$,
             $\alpha_{\rm CE}=3.0$, and $\gamma_{\rm CE}=1.5$, respectively. The
             dotted line represents a power-law function of form $f(t)\propto t^{\rm -1}$.}
 \label{single02}%
     \end{figure}

\section{Results for a normal DD channel}\label{sect:3}
\subsection{Birth rate for a single starburst}\label{subs:3.1}
In Fig. \ref{single02}, we show the evolution of the birthrates of
SNe Ia from DD systems for a single starburst with $Z=0.02$, where
the evolution is characterized by a DTD. For comparison, we {also
show a power-law function of the form $f(t)\propto t^{\rm -1}$ by
a dotted line in Fig. \ref{single02}. We can see from Fig.
\ref{single02} that the shape of the DTD depends on the treatment
of CE. Some recent measurements measure a similar power-law shape
with an index of $-1$ (Maoz \& Mannucci \cite{MAOZ11}). In Fig.
\ref{single02}, only when $\alpha_{\rm CE}=3.0$ ($\alpha_{\rm
CE}\lambda=1.5$), does the DTD follow a power-law shape of $t^{\rm
-1}$. If $\alpha_{\rm CE}\geq1.5$, we may obtain a power-law DTD,
while for $\alpha_{\rm CE}=1.0$ ($\alpha_{\rm CE}\lambda=0.5$),
not only does the shape not follows a power-law shape, but the
birth rate is also significantly lower than that for $\alpha_{\rm
CE}=3.0$ (see also Ruiter et al. \cite{RUITER09}). This is because
that a low $\alpha_{\rm CE}$ generally means a more heavy
shrinkage of the orbital separation to eject CE. The system is
then more likely to merge into a single star, not a DD system.
However, the birthrate is almost unaffected by variations in the
delay time in the $\gamma$-algorithm, i.e. the DTD does not follow
a power law. In the $\gamma$-algorithm, the separation of a DD
system surviving a CE phase is usually larger than that in the
$\alpha$-formalism (see Eq. \ref{gamma}), which leads to a very
long GWR timescale, and then a lower birth rate within 15 Gyr. It
is noteworthy to point out that the $\gamma$-algorithm favors the
production of SNe Ia with a long delay time (Meng et al.
\cite{MENGXC11b}). When $\alpha_{\rm CE}=3.0$, about two thirds of
the SNe Ia explode within 1 Gyr, which is qualitatively consistent
with observations (see the review of Maoz \& Mannucci
\cite{MAOZ11}), while in the other two cases, the SNe Ia older
than 1 Gyr are more common.

In Fig. \ref{single02}, there is a single spike at early times,
because the SNe Ia with delay times shorter than $10^{\rm 8}$ yr
are very rare for the DD channel. To save CPU time, we defined a
binary sample to be $10^{\rm 7}$ binaries. If we enlarge the
sample to be large enough, the single spike should disappear and
the line become continuous.

    \begin{figure}
    \centering
    \includegraphics[width=60mm,height=80mm,angle=270.0]{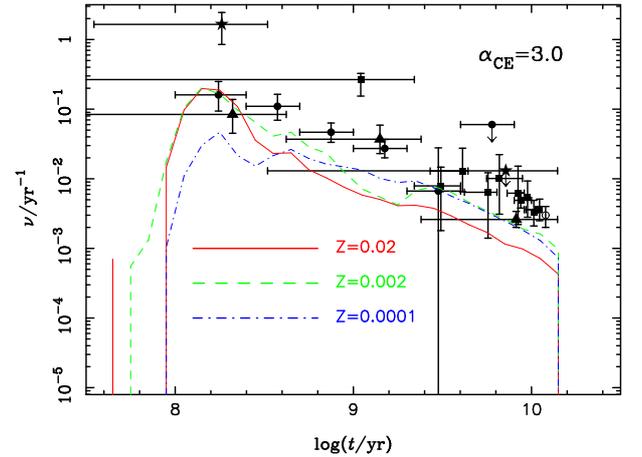}
    \caption{Evolution of the birthrates of SNe Ia for a single
             starburst of $10^{\rm 11}M_{\odot}$ with different
             metallicities,
             where $\alpha_{\rm CE}=3.0$. The circle is
             from
             Mannucci et al. (\cite{MAN05}), the filled circles are from Totani et al.
             (\cite{TOTANI08}), stars are from
             Maoz \& Badenes (\cite{MAOZ10a}), squares are from Maoz et al. (\cite{MAOZ10b}), and
             triangles are from  Maoz et al. (\cite{MAOZ11b}).}
 \label{dtdobv}%
     \end{figure}

We also check the dependence of the DTD shape on metallicity,
finding little significant influence using both the
$\alpha$-formalism and $\gamma$-algorithm. Fig. \ref{dtdobv} shows
the effect of metallicity on the DTD of $\alpha_{\rm CE}=3.0$. We
note that for the SNe Ia with the shortest delay time, a high
metallicity leads to a systematically earlier explosion time of
SNe Ia, which is mainly due to the effect of metallicity on the
stellar structure. For these SNe Ia, their delay time is dominated
by the evolutionary time of the primordial secondary and the
contribution of GWR is negligible. For a certain evolutionary
stage, a star with a high metallicity generally has a larger
radius, and may then fill its Roche lobe earlier. Hence, a given
binary system with a high metallicity usually enters the CE phase
earlier. If a DD system survives the CE evolution, an earlier
explosion is expected. Another feature in Fig. \ref{dtdobv} is
that low metallicity generally leads to a SN Ia with a long delay
time, especially for those older than $\sim1.5$ Gyr, which results
in a shallower power law at low metallicity. This also reflects
the effects of metallicity on the stellar evolution. Generally, a
star with a low metallicity usually produces a more massive white
dwarf, i.e. a less massive CE (Meng, Chen \& Han \cite{MENG08}).
After CE ejection, the surviving DD system of low metallicity has
a larger orbital separation than that of high metallicity.
Although a more massive WD also has a shorter GWR timescale, the
separation is the dominant factor when determining the GWR
timescale, and a large orbital separation then means a longer GWR
timescale.

In Fig. \ref{dtdobv}, we compare our power-law DTDs with recent
observations. Our results are marginally consistent with
observations.

    \begin{figure}
    \centering
    \includegraphics[width=60mm,height=80mm,angle=270.0]{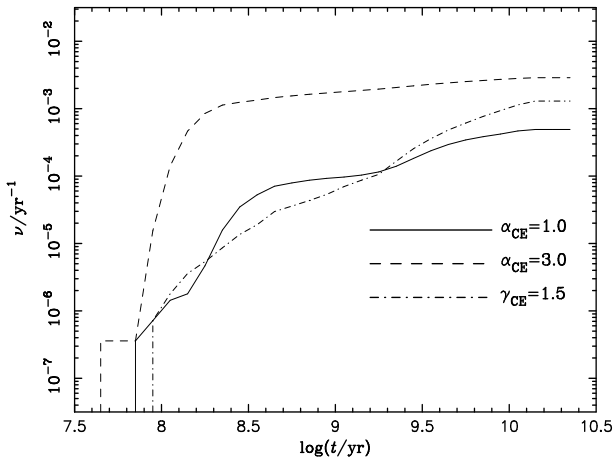}
    \caption{The evolution of the birth rates of SNe Ia for a constant
             star formation rate (Z=0.02, SFR=$5 M_{\rm \odot}{\rm yr^{\rm-1}}$).
             Solid, dashed, and dot-dashed lines show the cases of $\alpha_{\rm
             CE}=1.0$, $\alpha_{\rm CE}=3.0$, and $\gamma_{\rm CE}=1.5$, respectively. }
 \label{sfr02}%
     \end{figure}

    \begin{figure}
    \centering
    \includegraphics[width=60mm,height=80mm,angle=270.0]{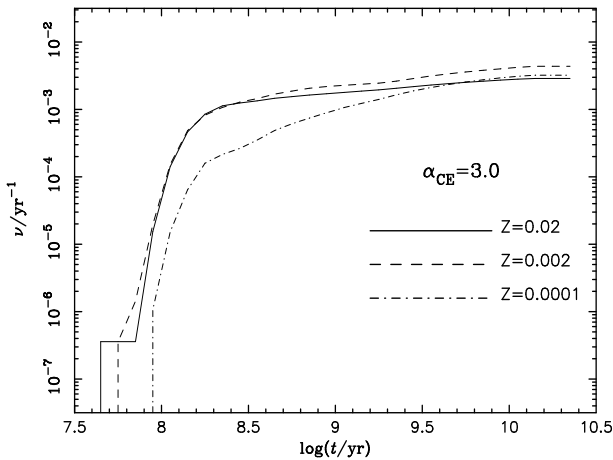}
    \caption{The evolution of the birth rates of SNe Ia for a constant
             star formation rate (SFR=$5 M_{\rm \odot}{\rm yr^{\rm-1}}$)
             with different metallicities, where $\alpha_{\rm CE}=3.0$.
             Solid, dashed, and dot-dashed lines show the cases of
             $Z=0.02$,$0.002$, and $0.0001$, respectively. }
 \label{sfrc}%
     \end{figure}

 \subsection{Birth rate for a constant star formation rate}\label{subs:3.2}
Fig. \ref{sfr02} shows the Galactic birth rates of SNe Ia (i.e.
$Z=0.02$ and SFR= 5.0$M_{\odot} {\rm yr}^{\rm -1}$) from the DD
channel. In the figure, the Galactic birth rate is around
0.5-3.0$\times10^{\rm -3}{\rm yr^{\rm -1}}$, which is marginally
consistent with that inferred from observations (3-7$\times10^{\rm
-3}{\rm yr^{\rm -1}}$, van den Bergh \& Tammann \cite{VAN91};
Cappellaro \& Turatto \cite{CT97}; Li et al. \cite{LIWD11}). For
the case of $\alpha_{\rm CE}=3.0$, the SNe Ia start at the age of
several tens of Myr and the birth rate increases remarkably until
about 0.2 Gyr. The birth rate then slowly increases, again because
SNe Ia with short delay times predominate the case of $\alpha_{\rm
CE}=3.0$. However, for the other two cases, the birthrates
permanently increase with delay time, owing to most of SNe Ia
having a long delay time.

Fig. \ref{sfrc} presents the effects of metallicity on the birth
rate of SNe Ia with a constant SFR of $5 M_{\rm \odot}{\rm
yr^{\rm-1}}$. From the figure, we can see that decreasing the
metallicity from $0.02$ to $0.0001$ may increase the present birth
rates by up to about 50\%, which is qualitatively consistent with
the discovery that the SN Ia rate in lower-metallicity galaxies is
higher than that in metal-rich environments ( Kister et al.
\cite{KISTLER11}). However, the present birth rate does not
monotonically increase with decreasing metallicity, i.e. the
highest value is for the case of $Z=0.002$. The reason for this
phenomenon is that explained in the above section, i.e. a low
metallicity may lead to a longer GWR timescale. If the GWR
timescale of DD system is so long that its delay time is longer
than 15 Gyr, it does not contribute to the present birth rate of
SNe Ia. Interestingly, the DD system was proposed to be the
progenitor of super-luminous SNe Ia (Astier et al.
\cite{ASTIER06}; Howell et al. \cite{HOW06}), which tend to
explode in metal-poor environments (Khan et al. \cite{KHAN11}).

\subsection{The normalization of DTD}\label{subs:3.3}
\begin{table}
\caption{DTD normalization($N_{\rm SN}/M_{\ast}$) from DD mergers
for different metallicities and different treatments
of CE in unit of $10^{\rm -3}/M_{\odot}$.}             
\label{table:1}      
\centering                          
\begin{tabular}{c c c c}        
\hline\hline                 
Z & $\alpha_{\rm CE}=1.0$ & $\alpha_{\rm CE}=3.0$ & $\gamma_{\rm CE}=1.5$ \\    
\hline                        
   0.02   & 0.098 & 0.58 & 0.26 \\      
   0.002  & 0.13  & 0.87 & 0.24 \\
   0.0001 & 0.22  & 0.64 & 0.29 \\
\hline                                   
\end{tabular}
\end{table}

Apart from the form of the DTD, there is also fairly good
agreement in terms of the DTD normalization $N_{\rm SN}/M_{\ast}$,
i.e. the time-integrated number of SNe Ia per stellar mass formed.
The range of its value is $(0.5-3.5)\times10^{\rm -3}/M_{\odot}$,
focusing on $2\times10^{\rm -3}/M_{\odot}$ (Maoz \& Mannucci
\cite{MAOZ11}). In Tab. \ref{table:1}, we present the DTD
normalization for DD mergers with different metallicities and
different treatments of CE evolution. The results here are again
marginally consistent with those from observations. The upper
limit to the DTD normalization here is only $0.87\times10^{\rm
-3}/M_{\odot}$, which is true for the case of $Z=0.002$ and
$\alpha_{\rm CE}=3.0$.

   \begin{figure}
   \centering
   \includegraphics[width=60mm,height=80mm,angle=270.0]{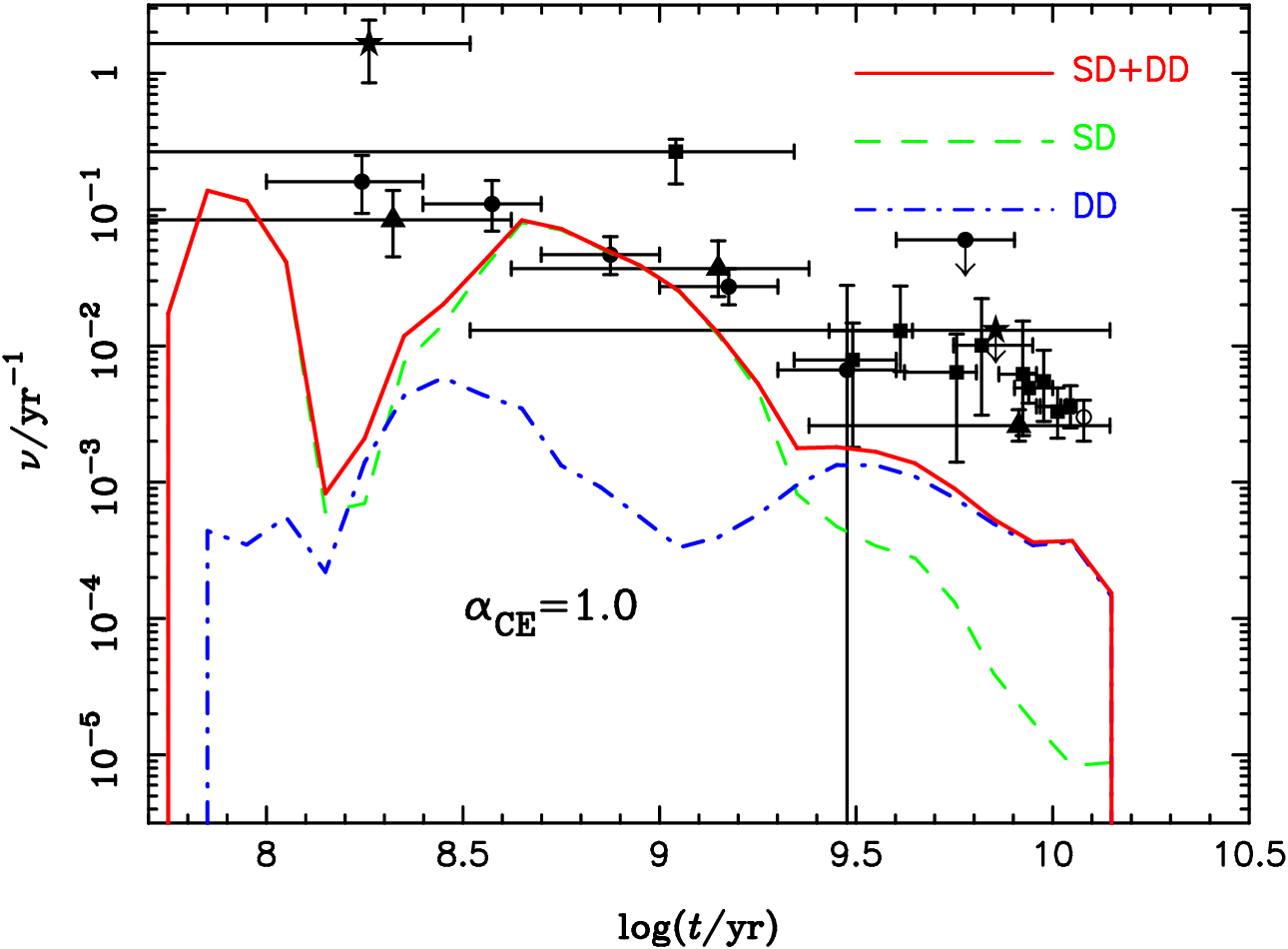}
   \includegraphics[width=60mm,height=80mm,angle=270.0]{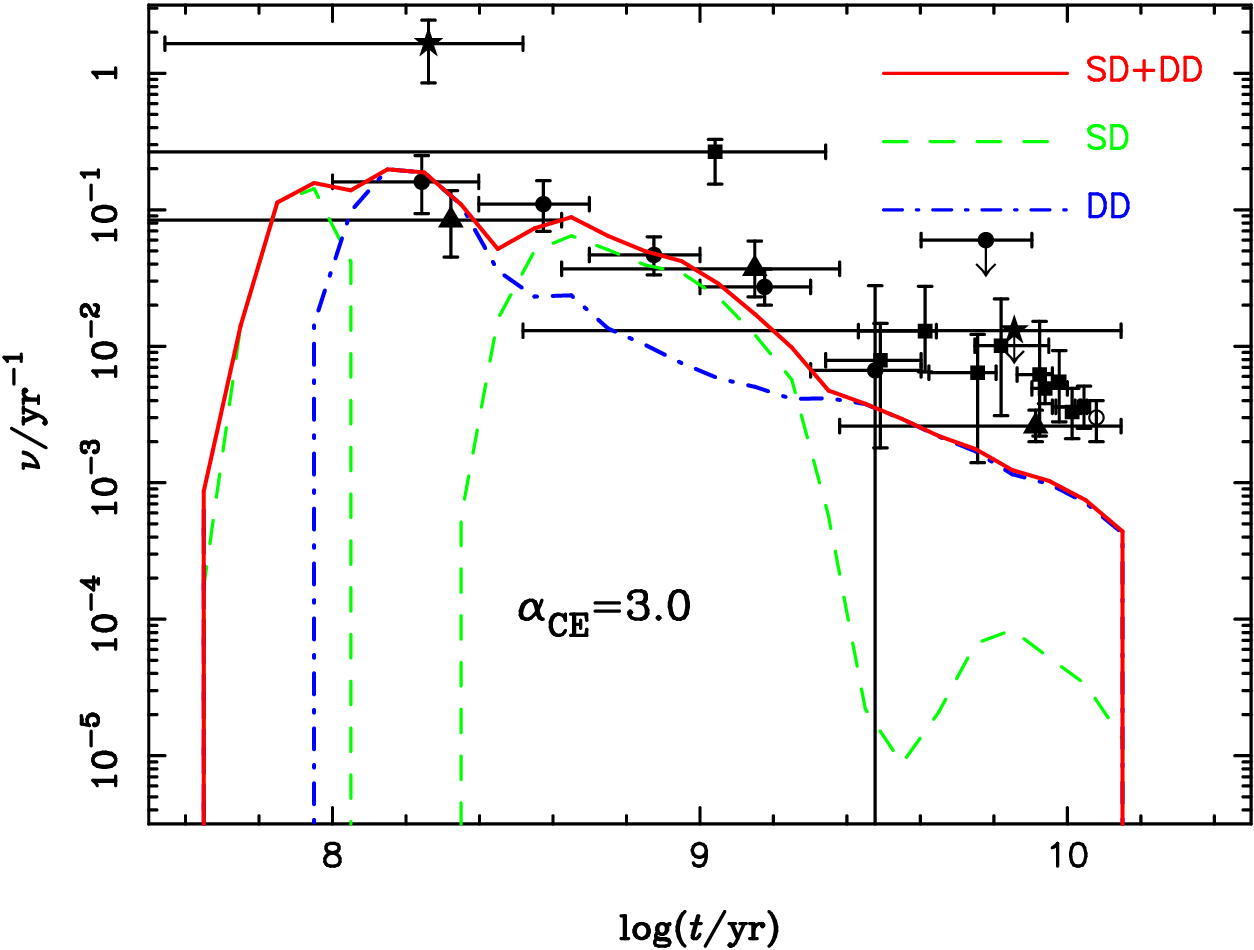}
   \caption{The evolution of the birth rate of SNe Ia for
   a single starburst of $10^{\rm 11} M_{\odot}$ and $Z=0.02$.
   Solid, dashed, and dot-dashed lines are from SD + DD, SD,
   and DD channels, respectively.
   The results for the SD channel are from Wang et al. (\cite{WANGB09b})
   and Meng \& Yang (\cite{MENGYANG10a})
   including the WD + He star, WD + MS, and WD + RG channels,
   while those for the DD channel are from this paper.
   The circle is from
   Mannucci et al. (\cite{MAN05}), the filled circles are from Totani et al.
   (\cite{TOTANI08}), stars are from
   Maoz \& Badenes (\cite{MAOZ10a}), squares are from Maoz et al. (\cite{MAOZ10b}), and
   triangles are from  Maoz et al. (\cite{MAOZ11b}).
   Top: $\alpha_{\rm CE}=1.0$; Botom: $\alpha_{\rm CE}=3.0$}
              \label{combination}%
    \end{figure}

\subsection{The DTD from SD + DD}\label{subs:3.4}
In Fig. \ref{combination}, we show a DTD obtained from a
combination of the SD and DD scenarios, where the SD scenario
includes the WD + He star, WD + MS, and WD + RG channels and the
mass transfers for these channels all occur via a stable RLOF
between a WD and helium/main sequence/redgiant star, not by wind
accretion (see Wang et al. \cite{WANGB09b} and Meng \& Yang
\cite{MENGYANG10a} in details). For the WD + He star channel in
Wang et al. (\cite{WANGB09b}), a helium star is in helium main
sequence or crosses the helium HG. As in the DD channel, the DTD
from the SD channel is also affected by the treatment of CE, so is
the  combined DTD. For $\alpha_{\rm CE}=1.0$, the SD channel
dominates the SNe Ia with a short delay time, while the DD channel
contributes most of the SNe Ia older than about 2 Gyr. The final
combined DTD is a weak bimodality, where the WD + He star channel
mainly contributes to SNe Ia with delay times shorter than 0.1
Gyr, WD + MS to those of $0.1-2$ Gyr and DD channel to those older
than 2 Gyr. At delay time shorter than 2 Gyr, this combined DTD is
not inconsistent with observations for large observational errors,
but the long delay-time part of the DTD is significantly lower
than observations by a factor up to 10. For $\alpha_{\rm CE}=3.0$,
the combined DTD is closely consistent with the observational data
for SNe Ia with delay times shorter than 2 Gyr, while for the SNe
Ia older than 2 Gyr, the combined DTD only marginally matches
observations. As in the case of $\alpha_{\rm CE}=1.0$, the DD
channel is also the dominant channel for the SNe Ia older than 2
Gyr. The contribution of the WD + RG channel could be neglected
(see also Lipunov et al. \cite{LIPUNOV11}). However, for young and
middle age SNe Ia, the contributions of the SD and DD scenarios
are comparable.

    \begin{figure}
    \centering
    \includegraphics[width=60mm,height=80mm,angle=270.0]{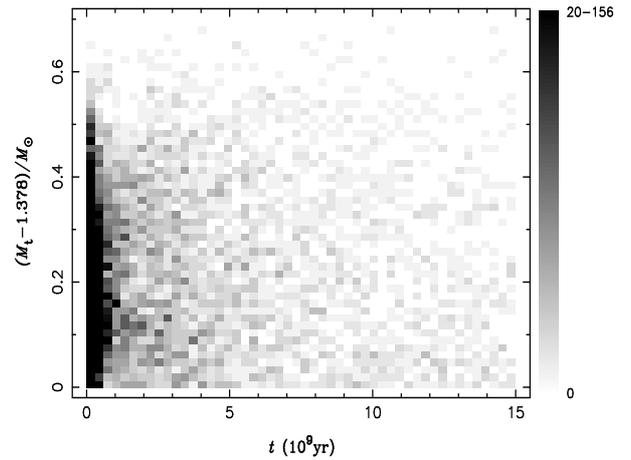}
    \caption{The evolution of the distribution of the total mass of
    the DD systems leading to SNe Ia for a
    single starburst, where Z=0.02 and $\alpha_{\rm CE}=3.0$.
    In the figure, all the cells move downward artificially
    by $1.378 M_{\odot}$.}
 \label{mtage}%
     \end{figure}

\subsection{The distribution of total mass}\label{subs:3.5}
Howell (\cite{HOWEL11}) suggested that the DD scenario has a
natural explanation for a relatively high SN luminosity in young
environments, namely that younger, more massive stars produce more
massive white dwarfs, which have more potential fuel than less
massive mergers, and then brighter SNe Ia (see also Maoz \&
Mannucci \cite{MAOZ11}). If this suggestion were right, we should
see a significant change in the total mass range of DD systems and
its average value, i.e. the range and intermediate value of the
total mass should decrease with delay time. In Fig. \ref{mtage},
we show the evolution of the distribution of the total mass, but
we do not find any significant expected evolution, i.e. the
distribution of the total mass is rather uniform across the whole
delay-time interval. The uniform distribution is produced mainly
because the dominant delay time for SNe Ia from DD systems is
determined by the GWR, not the evolutionary time of the secondary.

We investigated various physical properties when searching for a
correlation between delay time and a mass-dependent quality (the
masses of more massive WD $M_{\rm 1}$, less massive one $M_{\rm
2}$, mass ratio $M_{\rm 2}/M_{\rm 1}$, reduced mass $\frac{M_{\rm
1}M_{\rm 2}}{M_{\rm 1}+M_{\rm 2}}$, the radius of less massive WD,
the final separation when coalescence begins, and the total
angular momentum of the system at the moment of the onset of
coalescence). Despite our efforts, no correlation fulfills the
observational limit (the range of the parameter decreases with the
age of the SNe Ia, while the average value of the parameter
decreases/increases with the age) was found, which has roots in
the evolution of the total mass and mass ratio distributions.
Hence, under the framework of DD scenario, it may still be
difficult to explain the scatter in the brightness of the SNe Ia
or the Phillips relation and more efforts are needed to build a
bridge between the DD model and explosion model of SNe Ia.

In Fig. \ref{mtage}, most of the DD systems have a mass higher
than the Chandrasekhar mass by no more than 0.5 $M_{\odot}$.
During the merging process of a DD system, the less massive WD is
generally destroyed and its mass is accreted onto the more massive
WD. On the basis of Shen et al. (\cite{SHEN11}), we could expect
that after the destruction of the less massive WD, a giant-like
structure be produced during the accretion stage. Because of the
relatively long-lasting time that the giant-like structure
persists, we might expect to lose about half a solar mass from the
system (Soker \cite{SOKER11}). If this were a real scenario,
according to the results in Fig. \ref{mtage}, the final masses of
the most mergers could not reach the Chandrasekhar mass, and the
SNe Ia could not then be expected, even though the total mass of
the DD systems before merging exceeds the Chandrasekhar mass
limit.

    \begin{figure}
    \centering
    \includegraphics[width=60mm,height=80mm,angle=270.0]{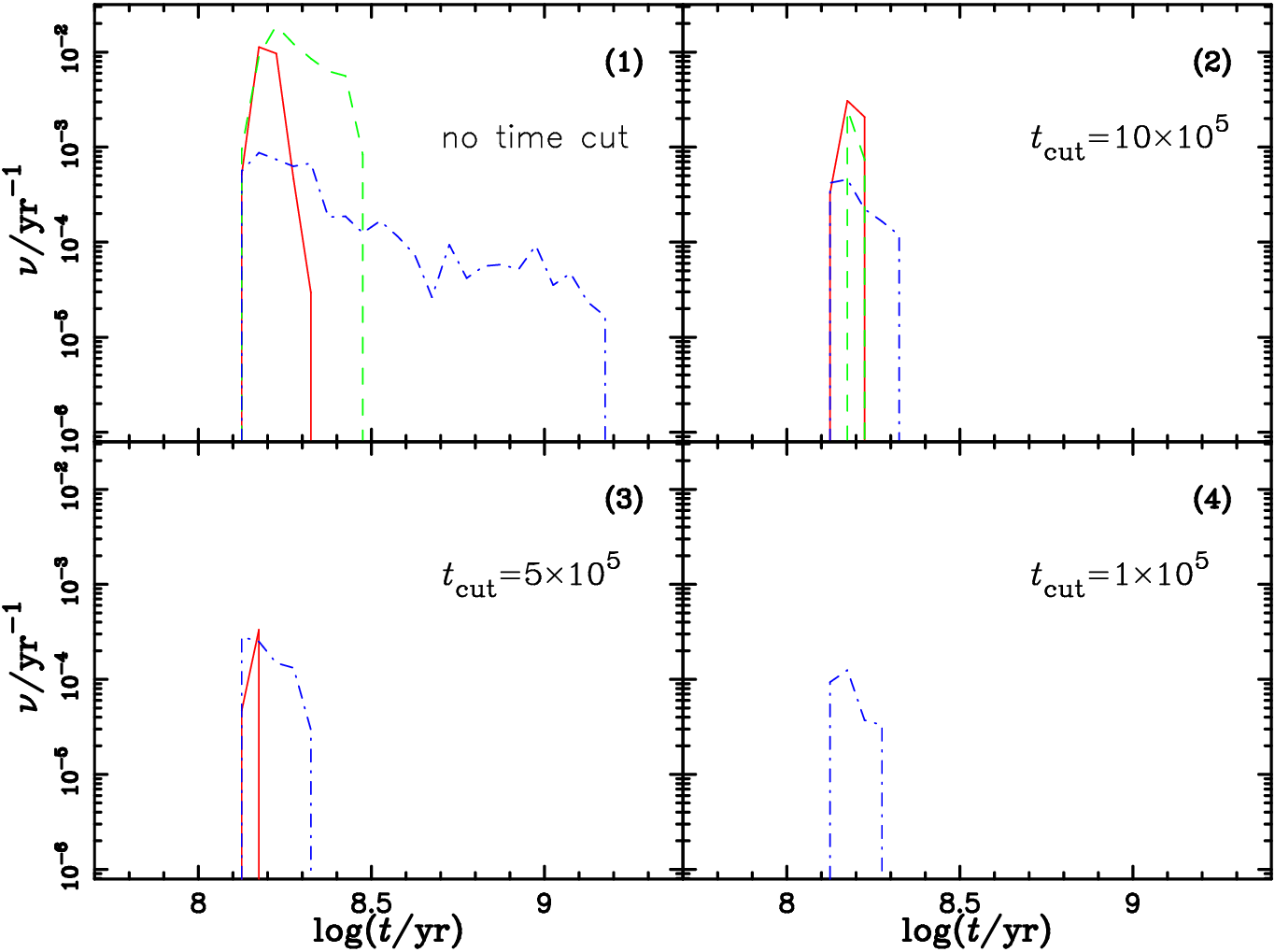}
    \includegraphics[width=60mm,height=80mm,angle=270.0]{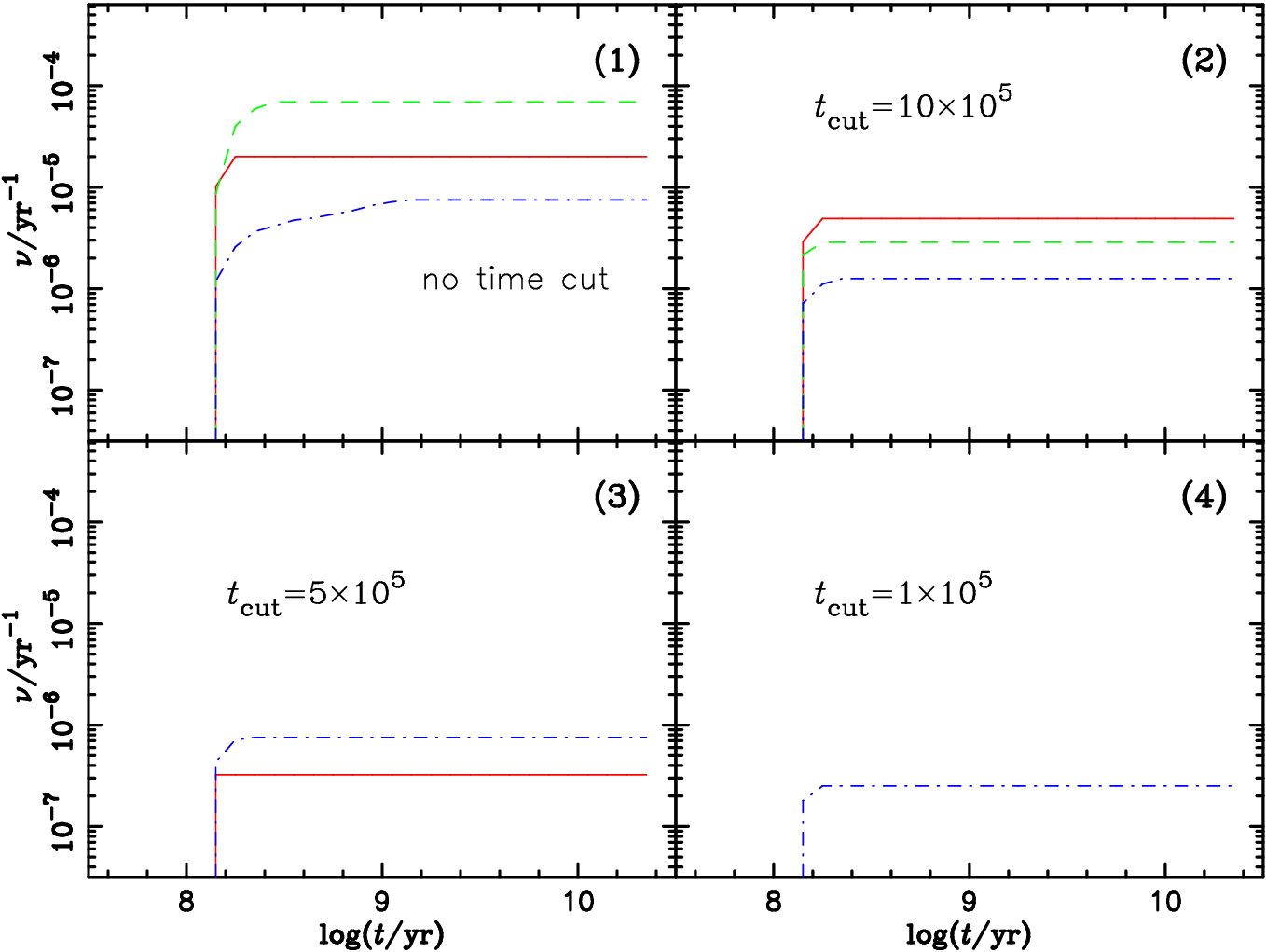}
    \caption{The evolution of the birth rate of SNe Ia from a CD channel
     for a single starburst of $10^{\rm 11} M_{\odot}$ (Top)
     and constant star formation rate of $5 M_{\rm \odot}{\rm yr^{\rm-1}}$ (Bottom) with different
     cutting times and different treatments of the CE,
     where $Z=0.02$. The delay time in the figure does not include the spinning-down time from
     magneto-dipole radiation. Solid, dashed, and dot-dashed lines are the cases
     of $\alpha_{\rm CE} =2.0$, $\alpha_{\rm CE}=3.0$, and $\gamma_{\rm CE}=1.5$,
     respectively.}
 \label{cdsingle}%
     \end{figure}

\section{The birth rate of SNe Ia from CD systems}\label{sect:4}
\subsection{Without spinning-down time}\label{subs:4.1}
In Fig. \ref{cdsingle}, we show the evolution of the birth rate of
SNe Ia from the CD channel for a single starburst and a constant
star formation rate with different treatments of CE, where the
spinning-down time from magneto-dipole radiation is assumed to be
insignificant and SN Ia occurs immediately after merging. The
cutting time means that only when a CD system merges within the
cutting time, is it assumed to explode as a SNe Ia. In the figure,
almost all of the SN Ia produced by CD systems have a delay time
shorter than $2\times10^{\rm 8}$ yr and the birth rate peaks at
$\sim 1.5\times10^{\rm 8}$ yr if we assume that all the potential
CD systems must merge within $10\times10^{\rm 5}$ yr after the CD
systems form, where the delay time is mainly determined by the
evolutionary time of the primordial secondary whose primordial
mass is generally higher than 3 $M_{\odot}$. Hence, if a
spinning-down time following magneto-dipole radiation did not play
a dominant role on the delay time of SNe Ia, the CD scenario may
only produce SNe Ia with very short delay times as expected by
Livio \& Riess (\cite{LR03}). In panel (1) of Fig. \ref{cdsingle},
some SNe Ia may have delay times as long as 1.5 Gyr, especially
for $\gamma_{\rm CE}=1.5$, which is from the ``pollution''
channel. Our results for cutting times shorter than
$10\times10^{\rm 5}$ yr are not polluted by the channel.

Even in the case of $t_{\rm cut}\leq10\times10^{\rm 5}$ yr, the
results are still significantly affected by the treatment of CE
evolution. For a $\alpha$-formalism, if $\alpha_{\rm CE}<2.0$, no
CD system survives the CE evolution. Even for $\alpha_{\rm
CE}\geq2.0$ ($\alpha_{\rm CE}\lambda\geq1.0$), the CD systems that
survived CE evolution will always merge within a timescale longer
than $1\times10^{\rm 5}$ yr. In contrast, the $\gamma$-algorithm
may produce CD systems with various merger timescales.

However, the peak value of the birth rate for the CD channel is
lower than that for the normal DD channel by $1-3$ magnitudes,
which means that the SNe Ia produced in the CD scenario are not
the main contributors to all SNe Ia. As shown in the bottom figure
of Fig. \ref{cdsingle}, no more than 1 in 1000 SNe Ia should be
produced by the CD scenario. The low birth rate results from the
constraint that the more massive hot core originates from the
primordial secondary.

    \begin{figure}
    \centering
    \includegraphics[width=60mm,height=80mm,angle=270.0]{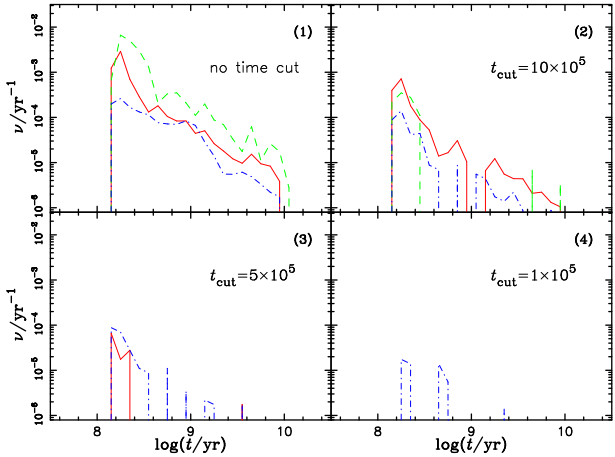}
    \caption{The evolution of the birth rate of SNe Ia from the CD channel
     for a single starburst of $10^{\rm 11} M_{\odot}$ with different
     cutting times and different treatments of CE,
     where $Z=0.02$. The delay time in the figure includes the spinning-down time from
     magneto-dipole radiation. Solid, dashed, and dot-dashed lines are the cases
     for $\alpha_{\rm CE} =2.0$, $\alpha_{\rm CE}=3.0$, and $\gamma_{\rm CE}=1.5$,
     respectively.}
 \label{cdbsingle}%
     \end{figure}

\subsection{With spinning-down time}\label{subs:4.2}
Ilkov \& Soker (\cite{ILKOV11}) suggested that the delay time for
SNe Ia produced by a CD system may be dominated by the spin-down
timescale related to magneto-dipole radiation torque, and the
timescale from an initial fast rotation $\Omega_{\rm 0}$ to a
critical angular velocity $\Omega_{\rm c}$ is
\begin{equation}
\begin{array}{lc}
\tau_{\rm B} \approx10^{\rm 8}\left(\frac{B}{10^{\rm 8} {\rm
G}}\right)^{\rm -2}\left(\frac{\Omega_{\rm c}}{0.7\Omega_{\rm
Kep}}\right)^{\rm
-2}\\
\hspace{0.75cm}\times\left(\frac{R}{4000 {\rm km}}\right)^{\rm
-1}\left(\frac{\sin\delta}{0.1}\right)^{\rm
-2}\left(\frac{\beta_{\rm
I}}{0.3}\right)\left[1-\left(\frac{\Omega_{\rm 0}}{\Omega_{\rm
c}}\right)^{\rm -2}\right] {\rm yr},
\end{array}
  \end{equation}
where $B$ is the magnetic field, $R$ is the radius of WD, and
$\beta_{\rm I}$ is a structural constant for the moment of
inertia. For simplicity, we assume that $\Omega_{\rm
0}=\Omega_{\rm Kep}$, $\Omega_{\rm c}=0.7\Omega_{\rm Kep}$,
$\beta_{\rm I}=0.3$, and $R=4000$ km. The spin-down timescale is
then
\begin{equation}
\tau_{\rm B}\approx5\times10^{\rm
7}\left(\frac{B\sin\delta}{10^{\rm 7} {\rm G}}\right)^{\rm -2}
{\rm yr},
\end{equation}
where $B\sin\delta$ follows a distribution of
 \begin{equation}
 \frac{dN}{d\log(B\sin\delta)}= constant,
 \end{equation}
for $10^{\rm 6} {\rm G}\leq B\sin\delta \leq 10^{\rm 8} {\rm G}$
(Ilkov \& Soker \cite{ILKOV11}). On the basis of this
distribution, we obtained the spin-down timescale using a Monte
Carlo simulations, and inferred the delay time of a SN Ia to be
the sum of the evolutionary timescale in forming a CD system, the
one for GWR, and the spin-down timescale if the total mass of the
CD system is lower than 1.48 $M_{\odot}$ otherwise the spin-down
timescale is neglected for those with total masses higher than
1.48 $M_{\odot}$ (Yoon \& Langer \cite{YOON04}).

In Fig. \ref{cdbsingle}, we show the DTD of SNe Ia produced by a
CD system, where the spin-down timescale is included. In the
figure, the DTD also follows a power law of $t^{-s}$, where $s$
ranges from 1.0 to 1.5 depending on the treatment of CE evolution.
This is a natural result since the distribution of the
$B\sin\delta$ is deduced from a power-law DTD (Ilkov \& Soker
\cite{ILKOV11}). Hence, the spin-down timescale is very important
to the SNe Ia produced by the CD channel since the power-law shape
does not occur according to the $\alpha$-formalism if the
spin-down timescale is not taken into account (see Fig.
\ref{cdsingle}). In addition, as expected by Kashi \& Soker
(\cite{KASHI11}) and Ilkov \& Soker (\cite{ILKOV11}), a few SNe Ia
produced by CD systems may have delay times as long as several
Gyr.

\section{Discussions and conclusions}\label{sect:5}
\subsection{Other possible channels for SNe Ia}\label{subs:5.1}
From our study, we have found that the birth rate of SNe Ia for DD
channels may only be marginally consistent with observations, and
the rate depends significantly on the treatments of CE. Only when
$\alpha_{\rm CE}\geq1.5$, may we obtain a DTD whose shape follows
a power law of $t^{\rm -1}$, and the birth rate is only consistent
with the lower limit of this trend derived from observations. For
the cases of $\alpha_{\rm CE}<1.5$ and $\gamma_{\rm CE}=1.5$,
neither the DTD shapes nor the birth rates match the available
observations. Badenes \& Maoz (\cite{BADENES12}) calculated the
merger rate of binary WDs in the Galactic disk based on
observational data in the Sloan Digital Sky Survey and concluded
that the merger rate of binary WDs with super-Chandrasekhar masses
would not control significantly the SNe Ia rate. Our results are
qualitatively consistent with the calculations of Badenes \& Maoz
(\cite{BADENES12}). If we consider the contribution of the SD
channel as done in Mennekens et al. (\cite{Mennekens10}), the
combination DTD may match observations for delay times $t<2.5$
Gyr, at least within observational errors, regardless of the value
of $\alpha_{\rm CE}$. However, when $t>2.5$ Gyr, the birth rate
from combination DTD is lower than that from observations.
Therefore, other channels or mechanisms may contribute to SNe Ia
with long delay times.

In this paper, the WD + He star, the WD + MS, and the WD + RG
channels are included for the SD channel, while the wind-accretion
channel is exclued. Chen, Han \& Tout (\cite{CHENXF11})
constructed a tidally-enhanced wind-accretion model where the
initial mass of RG donor may be as low as 1 $M_{\odot}$, i.e. the
delay time from the channel may be as long as 10 Gyr. On the basis
of Eq. (1) of Iben \& Tutukov (\cite{IBE84}), they obtained a
birth rate of $6.9\times10^{\rm -3} {\rm yr}^{\rm -1}$ from the
wind model. Although we believe that the birth rate is probably
overestimated since some parameter spaces considered to be valid
for SNe Ia production in Eq. (1) of Iben \& Tutukov (\cite{IBE84})
may not contribute to SNe Ia, the wind channel should improve our
results in this paper.

Some overluminous SNe Ia were observed and a super-Chandrasekhar
mass explosion was expected for these SNe Ia based on the amount
of $^{\rm 56}$Ni inferred (Astier et al. \cite{ASTIER06}; Howell
et al. \cite{HOW06}; Hicken et al. \cite{HICKEN07}; Scalzo et al.
\cite{SCALZO10}; Yuan et al. \cite{YUAN10}; Tanaka et al.
\cite{TANAKA10}; Yamanaka et al. \cite{YAMANAKA10}). These
overluminous SNe Ia are most likely to occur in metal-poor
environments (Khan et al. \cite{KHAN11}). Although they are
generally assumed to be produced by the coalescence of DD systems,
they may also originate in SD systems (Chen \& Li \cite{CHENWC09};
Liu et al. \cite{LIUWM10}; Hachisu et al. \cite{HKHN12}). However,
the super-Chandrasekhar mass explosion from the SD channel may
only contribute to SNe Ia with short delay times (Hachisu et al.
\cite{HKHN12}), and their contribution to all SNe Ia may be no
more than 0.3\% (Meng et al. \cite{MENGXC11b}).

\subsection{The effect of metallicity}\label{subs:5.2}
As noted in Sects. \ref{subs:3.1} and \ref{subs:3.2}, the birth
rate of SNe Ia depends on metallicity, especially for those with
long delay times. Meng et al. (\cite{MENGXC11}) found that
decreasing the metallicity may significantly increase the birth
rate of SNe Ia with long delay times, that are produced by the WD
+ RG channel, by a factor of about three. Some observations do
indeed find that a lower metallicity may correspond to a higher
SNe Ia birth rate (Kistler et al. \cite{KISTLER11}).

The picture emerging from some observations is remarkable, i.e.
most of these diverse DTD derived from different methods,
different environments, and different redshifts agree with each
other, in both form and absolute value (Maoz \& Mannucci
\cite{MAOZ11}; Graur et al. \cite{GRAUR11}). At delays of $t>1$
Gyr, a power law of index about $-1$ seems to be in little doubt,
although a index of $\sim-1.5$ could not be ruled out (Barbary et
al. \cite{BARBARY12}; Sand et al. \cite{SAND12}), while at delays
$t<1$ Gyr, the shape of the DTD might become either shallower or
steeper, or follow the same shape seen at long delays. In this
paper, we note that for the DD scenario, a low metallicity may
lead to a shallower DTD. Interestingly, for a SD channel, a
shallower DTD may also be obtained at low metallicity (Meng et al.
\cite{MENGXC11}). In addition, metallicity may also affect the
value of the birth rate of SNe Ia (Kistler et al.
\cite{KISTLER11}). Nevertheless, the level of the effect of
metallicity on the DTD might not be as high as we expect (see in
Fig. \ref{dtdobv}) since the DTDs from different environments
agree with each other. However, we still have to determine using
observations the precise effect of metallicity on the DTD since
metallicity does have an influence on the DTD, on both its form
and absolute value, and then the shape of the DTD at delays $t<1$
Gyr might become clear.

\subsection{The origin of the Phillips relation}\label{subs:5.3}
The Phillips relation is the most fundamental relation when a SN
Ia is used as a distance indicator, which implies that the
properties of SNe Ia are mainly dominated by one parameter.
However, the nature of this parameter remains unclear
(Podsiadlowski \cite{POD08}). Many efforts have been made to
resolve this problem, but most of these discussions have focused
on the Chandrasekhar mass model, in which the WDs explode as SNe
Ia when their masses are close to the Chandrasekhar mass limit. No
consensus has yet been reached at present. Here, we tried to
identify a parameter in the DD scenario to explain the scatter in
the luminosity of SNe Ia. Unluckily, no such a parameter was
found. In particular, the total mass of the DD systems is a poor
choice of parameter to illustrate the dependence of the average
luminosity of SNe Ia on their age. Hence, the origin of Phillips
relation still remains unclear.

\subsection{The birth rate of SNe Ia from CD channel}\label{subs:5.4}
We have found that the upper limit to the contribution of the SNe
Ia produced by the CD channel to the total SNe Ia is only about
0.1\%. One may argue that this result may seriously depend on the
treatment of CE evolution since the treatment here is very simple
and the real scenario may be very complex (Kashi \& Soker
\cite{KASHI11}; Passy et al. \cite{PASSY12}). At present, it
remains very difficult to construct a more complete model in a BPS
study. However, since the simple treatment may succeed in
reproducing the forms of many special objects, such as planetary
nebulae and subdwarf B stars (Han et al.
\cite{HAN95,HAN02,HAN07}), the results here could be reasonable to
at least first order.

However, one may remark that since the CD systems had hardly
survived the CE evolution, we might ignore the systems that almost
survive the CE evolution, especially for a lower $\alpha_{\rm
CE}$. These systems might also explode if their spinning-down time
is long enough to permit the remaining CE material to be lost by a
wind. Because the results of 0.1\% is derived from the
$\alpha$-formalism of $\alpha_{\rm CE}\geq2.0$, we might
underestimate the birth rate of SNe Ia predicted by the CD
scenario. However, if $\alpha_{\rm CE}<2.0$, no CD system can
survive the CE evolution. This is a natural result since a low
$\alpha_{\rm CE}$ means that a system needs to shrink its
separation more dramatically and release more orbital energy to
eject the CE. We do not verify whether this merger occurs because
this would require consideration of the details of CE evolution
and many assumptions have to be made, which may lead to too many
uncertainties. Fortunately, our results place a constraint on the
problem. For a high $\alpha_{\rm CE}$, the systems that merge
under the situation of low $\alpha_{\rm CE}$ could survive, while
those surviving the CE evolution with low $\alpha_{\rm CE}$ may
lead to a larger orbital separation. Hence, the panel (1) in Figs.
\ref{cdsingle} and \ref{cdbsingle} should give a safe upper limit
on the birth rate of SNe Ia based on the CD scenario, where all
the potential CD systems should be included. Even so, the total
contribution from all the potential CD systems to all SNe Ia is no
more than 1\%. In addition, the spin-down mechanism is very
important to the DTD shape of SNe Ia derived for the CD systems,
especially for the $\alpha$-formalism. We cannot obtain a
power-law form if the spin-down timescale due to the
magneto-dipole radiation torque is not taken into account.
\\
\\
In summary, we have calculated the evolution of the birth rate of
SNe Ia from DD and CD channels. We found that the treatment of the
CE evolution has a great influence on the final DTD shape, and
only for the $\alpha$-formalism when $\alpha_{\rm CE}\geq1.5$, the
shape of the DTD is consistent with that derived from
observations, but the birth rate of this case only marginally
matches that derived observationally. For a $\alpha$-formalism
with a lower $\alpha_{\rm CE}$ and the $\gamma$-algorithm, the
birth rate of SNe Ia from DD systems is much lower than those
derived from observations, and the shape of the predicted DTD does
not follow a power law. Metallicity has  almost no influence on
the shape of DTD except in terms of the index of the power law,
but may increase the value of the birth rate of SNe Ia. When the
SD channel including WD + He star, WD + MS, and WD + RG channels
is incorporated, the theoretical DTD matches the observations very
well for SNe Ia younger than 2.5 Gyr, while for those older than
2.5 Gyr, the theoretical birth rate is slightly lower than that
derived from observations. We also need to incorporate other
channels or mechanisms that contribute to the SNe Ia explosion. As
suggested by Soker (\cite{SOKER11}) and Ilkov \& Soker
(\cite{ILKOV11}), the CD scenario is a possible channel
contributing to SNe Ia and we have found three potential channels
that may produce the CD systems. However, the strict upper limit
to the contribution from the CD scenario to all SNe Ia is 1\%.

\begin{acknowledgements}
We are very grateful to the anonymous referee for his/her
constructive suggestions, which improved the manuscript greatly.
This work was partly supported by the Natural Science Foundation
of China (11003003), the Project of Science and Technology from
the Ministry of Education (211102), and the Key Laboratory for the
Structure and Evolution of Celestial Objects, Chinese Academy of
Sciences.
\end{acknowledgements}

\newpage

\end{document}